\documentclass[10pt,a4paper,superscriptaddress,aps,prd,showkeys,showpacs,nofootinbib,reprint]{revtex4-2}
\usepackage[utf8]{inputenc}
\usepackage{verbatim}
\usepackage{amsmath,amsfonts,amssymb}
\usepackage[breaklinks,colorlinks]{hyperref}
\usepackage{natbib}
\usepackage{tikz}
\usepackage{float}
\usepackage{graphicx}
\usepackage{adjustbox}
\usepackage{tabularx}
\usepackage{amsbsy}
\usepackage{braket}
\usepackage[singlelinecheck=off,justification=raggedright]{caption}
\usepackage{subcaption}
\hypersetup{%
,urlcolor=blue
,citecolor=blue
,linkcolor=blue
}

\begin{document}

\title{Stable cosmic vortons in bosonic field theory}
%\newline
%\normaltext{- the importance of plasma ray tracing - }

\author{R.~A.~Battye}
\email[]{richard.battye@manchester.ac.uk}
\affiliation{%
Jodrell Bank Centre for Astrophysics, School of Natural Sciences, Department of Physics and Astronomy, University of Manchester, Manchester, M13 9PL, U.K.
}

\author{S.~J.~Cotterill}
\email[]{steven.cotterill@postgrad.manchester.ac.uk}
\affiliation{%
Jodrell Bank Centre for Astrophysics, School of Natural Sciences, Department of Physics and Astronomy, University of Manchester, Manchester, M13 9PL, U.K.
}

\label{firstpage}

\date{\today}

\begin{abstract}
Stable ring solutions supported by the angular momentum caused by superconducting charge and current have been suggested to exist in the gauged $U(1)\times U(1)$ field theory. We construct potentially cosmologically relevant solutions using gradient flow for the first time and present the strongest evidence to date that they are stable to axial and, more importantly, non-axial perturbations. More importantly, we illustrate quantitative agreement with semi-analytic predictions based on the thin string approximation which validates worldsheet action approaches to their formation and evolution.
\end{abstract}

\keywords{ll}

\maketitle

{\it Introduction: }Cosmic strings will be formed in a cosmological phase transition where the vacuum manifold can support non-contractible loops. They have been studied in a wide range of cosmological contexts, see \cite{V&Sbook} for a detailed review. 

It was suggested by Witten~\cite{Witten1985} that they could act as superconducting wires with interesting consequences. He illustrated this point using a gauged $U(1)\times U(1)$ bosonic field theory\footnote{He discussed the possibility of fermionic string superconductivity which will not be discussed here.} but for our purposes it will be sufficient to consider the neutral limit~\cite{Peter1992} where only the vortex field, $\phi$, is gauged with Lagrangian density
\begin{eqnarray}
    \mathcal{L} = (\mathcal{D}_\mu\phi)(\mathcal{D}^\mu\phi)^* + \partial_\mu\sigma\partial^\mu\sigma^* - \frac{1}{4}F_{\mu\nu}F^{\mu\nu} \nonumber\\- \frac{\lambda_\phi}{4}(|\phi|^2-\eta_\phi^2)^2 - \frac{\lambda_\sigma}{4}(|\sigma|^2 - \eta_\sigma^2)^2 - \beta|\phi|^2|\sigma|^2 ,
    \label{lag}
\end{eqnarray}
where $\mathcal{D}_\mu = \partial_\mu - igA_\mu$, $F_{\mu\nu} = \partial_\mu A_\nu - \partial_\nu A_\mu$. The parameters ${\cal P} =(\eta_\phi,\eta_\sigma, \lambda_\phi,\lambda_\sigma,\beta,g)$ are all real positive constants which will be treated as dimensionless, imposing units on the space-time coordinates and the energy. We will choose parameters so that one of the $U(1)$ fields, $\phi$, is spontaneously broken allowing for vortex solutions, and the other, $\sigma$, remains unbroken with current 
\begin{equation}
 {\bf j}={1\over 2i}\left(\sigma^*\mathbf{\nabla}\sigma-\sigma\mathbf{\nabla}\sigma^*\right)\,,
\end{equation}
and conserved charge,
\begin{equation}
    Q={1\over 2i}\int d^3{\bf x}\left(\sigma^*{\dot{\sigma}}-\sigma{\dot\sigma^*}\right)\,.
\end{equation}

In what follows we will use the specific choices ${\cal P
}=(1,0.61,1,10,3,\frac{\sqrt{0.5}}{2})$ which have been chosen to allow superconducting vortex solutions in cylindrical polar coordinates $(\rho,\theta,z)$ with vortex solution $\phi(\rho,\theta)=|\phi|(\rho)\exp[i\theta]\,,$ and a  condensate $\sigma(\rho,z,t)=|\sigma|(\rho)\exp[i(\omega t+kz)]$.
The constants $\omega$ and $k$ are associated with the conserved charge, $Q$, and current. Within this ansatz, their impact can be absorbed into a re-definition of the potential $V\rightarrow V-\chi|\sigma|^2$ where $\chi=\omega^2-k^2$. Solutions with $\chi<0$ are referred to as magnetic, whereas those with $\chi>0$ are electric and $\chi=0$ are chiral. See \cite{V&Sbook} for a  detailed discussion of superconducting string solutions and choice of parameters necessary to achieve this; suffice to say there are a wide range parameters which will lead to similar results to those presented here. We note that it has been suggested that cosmic strings are generically current carying~\cite{Davis:1995kk}.

Davis and Shellard~\cite{Davis1988b} suggested that superconducting strings carrying both current and charge, bent round into a ring called a vorton, could be produced during the evolution of the Universe. If they were absolutely stable they could either come to overclose the universe, since they would scale like matter, or have a number of interesting cosmological effects~\cite{Ostriker1986,Brandenberger1996,Carter:1999an}, including being some or all of the dark matter of the Universe~\cite{Davis1989,Martins1998a,Martins1998b,Cordero-Cid:2002hmv,Peter:2013jj,Auclair:2020wse}.

{\it Thin sting approximation:} The vortons are likely to be stable to radial perturbations since the natural propensity for string loops to reduce in size due to their relativistic tension is balanced by the centrifugal effects of the angular momentum which is caused by the current flowing around the loop. One can model this by treating the string as a line-like defect in the ``thin string approximation" (TSA). For a loop of length $L=2\pi R$ constructed from a thin string solution with condensate profile $|\sigma|(\rho)$ the energy can be estimated as 
\begin{equation}
    E=\left(\mu-{1\over 4}\lambda_{\sigma}\Sigma_4\right)L+{2Q^2\over\Sigma_2L}\,,
\end{equation}
where $Q=2\pi\omega L$, $kL=2\pi N$, $N$ is the winding  of the current and 
\begin{equation}
    \Sigma_n=2\pi\int\rho|\sigma|^nd\rho\,.
\end{equation}
One can minimise the energy with respect to $L$ to obtain a prediction for the vorton radius, $R$, which we will test using field theory simulations later on.

In addition it is possible~\cite{Martin1994,Carter1989b,Carter1993} to investigate the stability of these ring solutions to transverse and longitudinal perturbations  by supplementing the Nambu action with fields representing the charge and current on the worldsheet - ``worldsheet action approaches". In particular, one can investigate the impact of small perturbations to the position of the string $\propto \exp [i(\Omega_m t-m\theta)]$ where the $\Omega_m=2\pi f_m$ is  predicted and will be tested in what follows.

{\it Previous attempts to construct stable vortons in field theory: }To date there has been no explicit demonstration of the existence and stability of vorton solutions in \eqref{lag}, but they are often assumed in cosmological applications.  A number of previous works have made some progress.
\begin{itemize}
    \item The interaction term was modified to $\beta^{\prime}|\phi|^6|\sigma|^2$~\cite{Lemperiere2003b} in order to strengthen the trapping potential for the condensate and the stability of vortons to elliptical perturbations were studied with $g=0$, that is, for a global symmetry.
     \item Global vortons were constructed in \eqref{lag} with  ${\cal P}=(1,1,3,2,2,0)$~\cite{Battye2009a}, but they were shown to be unstable in 3D field simulations to a mode with $m=4$, thought to be sourced by the numerical discretization. The characteristics of these vortons and their stability properties were not predicted by the TSA.
    \item Gauged vorton solutions with $N=1$ have been constructed~\cite{Garaud2013}. While this work shows that solutions similar to vortons can exist\footnote{In fact, these structures are much closer to those found in two-component Bose-Einstein condensates~\cite{Battye2002,Metlitski_2004}, albeit with a gauge symmetry.}, they are not cosmologically relevant since their size is very small, and in fact length-scales associated with the fields are of order the size of the overall defect. In contrast, vortons of cosmological importance should be expected to have radii that are many orders of magnitude larger than the core width.
    \item Kinky vortons~\cite{Battye2008}, a 2D analogue of vortons have been constructed, aided by an exact analytic solution for the superconducting domain wall solution, and their stability was confirmed using numerical and analytic techniques based on the TSA~\cite{Battye2009b}.
\end{itemize}
The objective of the work presented in this {\it letter} is to apply the method used to construct global vortons~\cite{Battye2009a}, and use the TSA to predict their properties and study their stability as was done for kinky vortons~\cite{Battye2009b}. We will not use a modified interaction term and the vortons we will produce have $N\gg 1$, much more indicative of the cosmologically relevant regime and around the largest possible in any numerical simulation with current technology.

{\it Constructing vorton solutions using gradient flow: }We have used numerical methods, similar to those used in \cite{Battye2009a}, to construct vorton solutions for given values of $Q$ and $N$ for a given set of parameters by imposing axial symmetry and using the cartoon method~\cite{Alcubierre1999}. The addition of the gauge fields requires the numerical methods to use techniques that will later allow the solutions to be evolved under the equations of motion while maintaining the gauge fixing condition throughout~\cite{Creutz1983}.

\begin{figure}[t]
    \centering
    \begin{subfigure}{\linewidth}
        \centering
        \includegraphics[width=\linewidth,trim={20cm 3cm 22cm 9cm},clip]{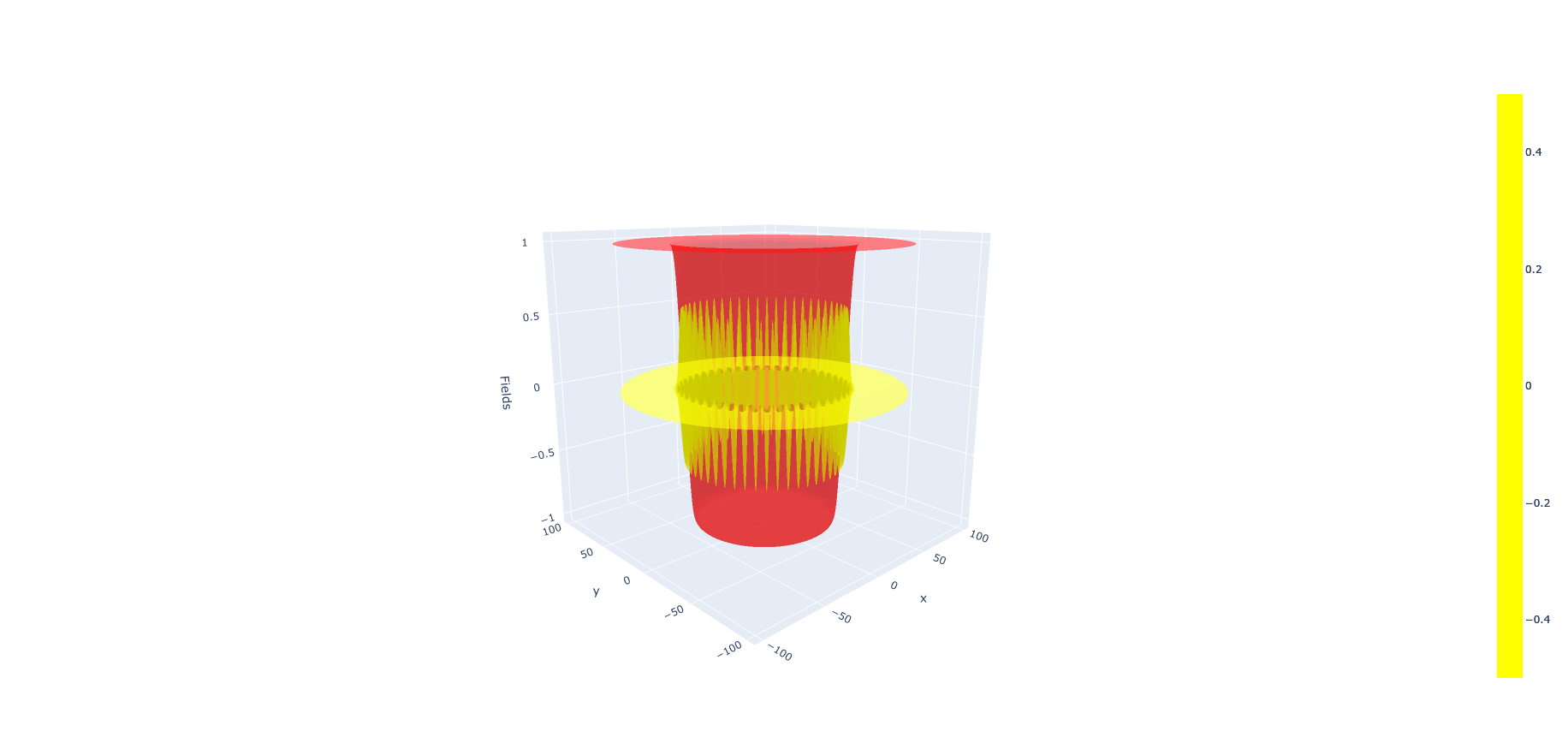}
    \end{subfigure}
    \begin{subfigure}{\linewidth}
        \includegraphics[width=0.9\linewidth]{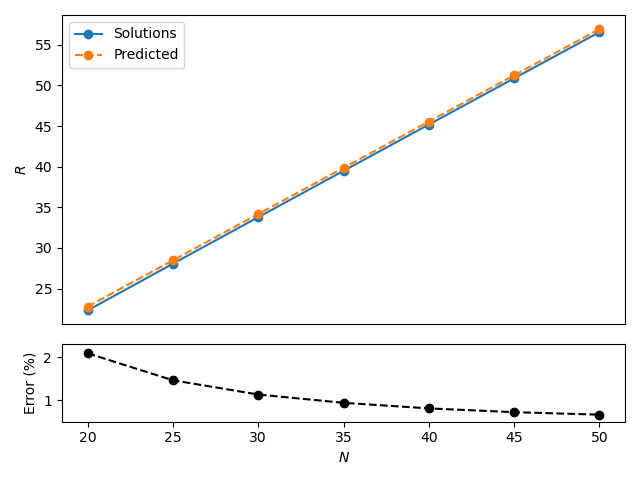}
    \end{subfigure}
    \caption{(top) The field profiles in the $z=0$ plane, with the red surface showing $\text{Re}(\phi)$ and the yellow surface showing $\text{Re}(\sigma)$. In this plane, $\text{Im}(\phi)=0$ and $\text{Im}(\sigma)$ looks similar to $\text{Re}(\sigma)$, acting to keep $|\sigma|$ axisymmetric. The winding of the condensate around the vorton and localisation to the string core, $\phi=0$, are clearly displayed. (bottom) The radii of vorton solutions as a function of $N$ (keeping $Q/N=31.89$ fixed) compared with the predictions from the TSA.}
    \label{fig:solution}
\end{figure}

\begin{figure}[t]
    \centering
    \includegraphics[width = \linewidth,trim={0 0 1cm 1cm},clip]{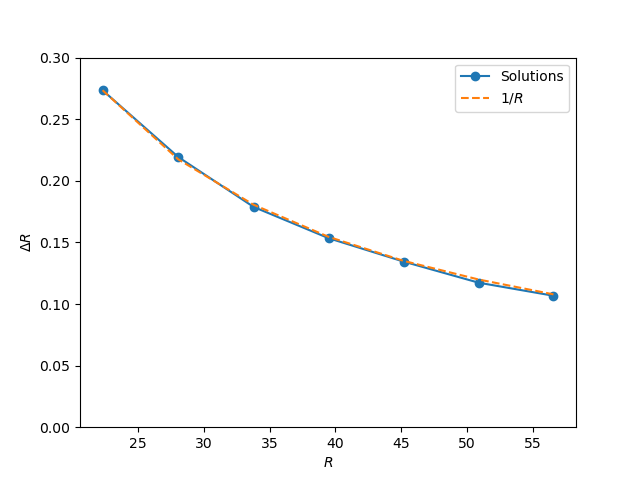}
    \caption{The difference between the radius of the vorton as measured by the core of the string - the measure that we use in the rest of the paper and for the x-axis of this plot - and the radius measured by the peak of the condensate - the larger of the two. The curve clearly exhibits a $1/R$ shape, suggesting that it is caused by curvature effects and will be negligible in cosmological applications.}
    \label{fig: deltaR}
\end{figure}

\begin{figure*}[t]
    \centering
    \includegraphics[width=\textwidth]{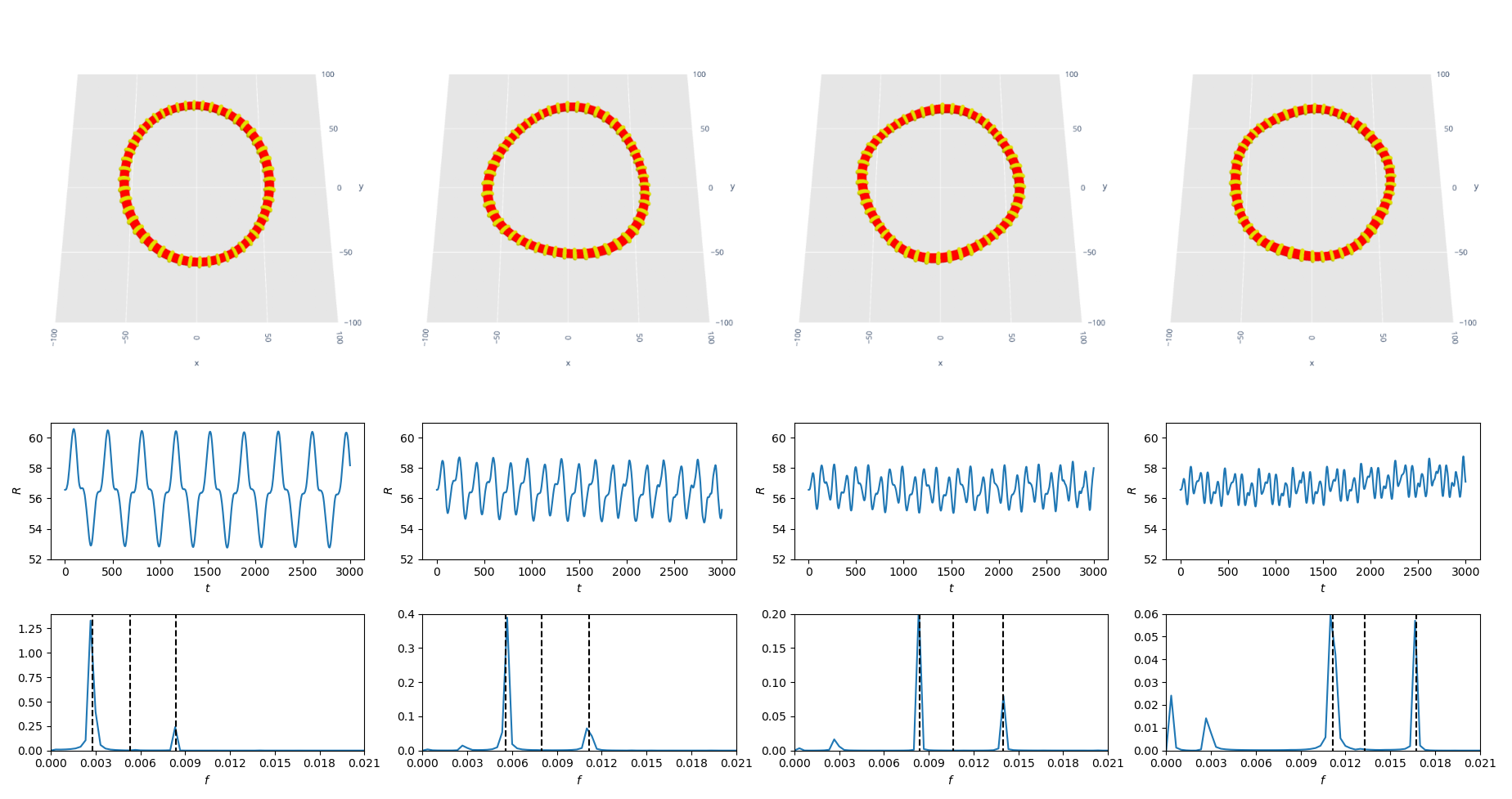}
    \caption{Simulations of perturbed vortons with 3D numerical field theory. On the top row, we show isosurfaces of the fields at $t=1000$, with $|\phi|=\frac{3}{5}\eta_\phi$ shown in red and $\text{Re}(\sigma)=\frac{1}{10}\eta_\sigma$ shown in yellow, demonstrating the distortions to the shape of the vorton caused by the $m=2-5$ (left to right) perturbations. The central plots show the position of the string over time in the $y=0$ plane (on the $x>0$ side) and the bottom row of images display the Fourier transforms, with the black dotted lines indicating the predicted frequencies of oscillation from the TSA. Two of the predicted frequencies are seen, and the third is expected to correspond to an almost entirely longitudinal oscillation for close to chiral vortons, and therefore not be apparent in positional data.}
\label{fig:spectra}
\end{figure*}

In Fig.~\ref{fig:solution} we present the fields for a solution with $Q=1594.3$ and $N=50$ - which we will use in subsequent investigations of stability - and the radius, $R$, as a function of $N$ keeping $Q/N=31.89$ fixed. The calculated radius is compared to that predicted from the TSA and we find good agreement between the two with $\sim 1\%$ accuracy for $N\gtrsim 40$. We find that $R\propto N$ for $Q/N$ fixed as is predicted by the TSA. When compared to the field profiles predicted by the TSA for the same set of parameters, the numerically generated vortons show some differences. Notably, we show in figure \ref{fig: deltaR} that the string cores in $\phi$ and maximum of the condensate in $\sigma$ do not exactly coincide as would be predicted in the TSA. However, we see that the offset reduces as the radius increases and are seen to tend to zero like $1/R$ indicating that they are due to curvature effects not included in the TSA. We note that cosmologically relevant vortons will typically have much larger radii than those presented here and we would expect these effects to be negligible there suggesting that the TSA can be used to predict the properties of cosmological vortons. We have found similar results for other parameter sets and, although this sweep is by no means exhaustive, we see this as strong quantitative confirmation of the TSA.

{\it 3D field theory simulations of vortons and their stability: }We have investigated the stability of the numerically generated vortons by interpolating them on to a Cartesian grid and evolving the equations of motion, initially imposing axial symmetry and then relaxing this assumption to evolve the full 3D equations of motion with a grid of $401^3$ points. During the evolution it is necessary to maintain the gauge condition to high precision, which we measure by calculating the absolute value of the deviation from the gauge condition, and averaging it at each timestep. Although this quantity does slowly grow in all of our simulations, it remains very small throughout - always below $10^{-3}$.

We were able to perturb the vorton radially by using a modified value of $\omega\approx 0.788$, a $10$\% reduction from the original value of $\omega\approx 0.875$ for the stationary solution. As expected the solution is stable to radial perturbations and the frequency of oscillation around the minimum is predicted to be $f_0^{\rm TSA}\approx 2.79\times 10^{-3}$. By Fourier-analyzing $R(t)$ we are able to numerically measure $f_0^{\rm num}\approx 2.83\times 10^{-3}$ with $\Delta f= 1.67\times 10^{-5}$ due to the large dynamic range ($t=6\times 10^4$) of the simulation. Again we see good agreement between the numerical results and the predictions from the TSA.

The global vortons studied in \cite{Battye2009a} were found to be unstable to non-axial $m=4$ perturbations sourced by the numerical discretization and there it was only possible to evolve the vortons for a short period of time. In the case studied here, the TSA predicts that vortons with $-4\times 10^{-4}<\chi<8\times 10^{-4}$, close to chiral, are stable to non-axial perturbations for all $m$. The vorton we have been using has $\chi\approx 2\times 10^{-4}$, in the electric regime, and is predicted to be stable. We have evolved the interpolated solution up to $t=10000$ - a value only set by reasonable use of computer resources - which corresponds to around 28 full rotations of the current around the ring. No obvious instability is seen ($<2\%$ variations in the radius) and Fourier analysis of $R(t)$ yields $f_0^{\rm num}=2.8\times 10^{-3}$. This evolution is longer than any other claimed numerical evolution and again we see quantitative agreement with the TSA.

We have perturbed the radius of the vorton by making the modification $\sigma\rightarrow\sigma(1+\epsilon\sin m\theta)$ with $\epsilon=0.1$. We find that this excites modes predicted by the TSA with frequency $f_m$. We evolve the solutions for $m=2-5$ up to $t=3000$ and plot a snapshot of the fields at $t=1000$, $R(t)$ and Fourier analysis of $R(t)$ in Fig.~\ref{fig:spectra}. The snapshots exhibit intuitive perturbations of the fields and the plots of $R(t)$ show clearly bounded oscillations expected for a stable vorton. The Fourier analysis of $R(t)$ exhibits excellent quantitative agreement between the spectra and the predictions of the TSA albeit with slightly worse resolution than in the axial case due to the shorter simulation time. Two of the frequencies predicted by the thin string approximation are excited by the perturbations we have applied. The vorton which we are studying is close to chiral and in such a case one of the predicted frequencies is expected to correspond to an oscillation that is almost entirely longitudinal, while the other two are a mix of longitudinal and transverse. The third mode is therefore not detected in $R(t)$ because it only tracks transverse oscillations.

{\it Conclusions: }Previous to our work there was some evidence that vortons might exist in field theory, but we believe our results represent a significant strengthening of that evidence implying a lower bound on the timescale for any instabilities which might be due to complicated non-linear interactions in the core of the superconducting vortices. For the case we have studied we have found solutions whose physical characteristics agree with the TSA. The solutions were shown to be stable to non-axial perturbation within 3D numerical field theory simulations and we were able to measure perturbed frequencies which again are seen to be compatible with those predicted by the TSA.

We have explored, but not reported in detail, some other parameter sets where we find similar results. We will discuss our full investigation, and a number of other interesting features we have observed, in a more detailed study that will be presented soon. It should also be noted that vorton-like solutions can, in principle, be formed in condensed matter systems~\cite{Battye2002,Metlitski_2004} but that, to our knowledge, this has not yet been attempted in an experimental context.

The most powerful implication of our study is that one can reliably use the TSA and beyond Nambu, worldsheet action approaches to vorton dynamics~\cite{Carter:2011ab} to make cosmological predictions for their evolution. It is far from simple, of course, to do this, but it is now on a firm footing, and something which should be explored more fully.

{\it Acknowledgements: }We would like to thank Jonathan Pearson for collaboration in the early stages of the work presented in this paper. We thank Paul Sutcliffe and Paul Shellard for comments on the text.

\bibliographystyle{apsrev4-1}
\bibliography{refs.bib}

\end{document}